\documentclass[12pt]{iopart}

\usepackage{iopams} 

\usepackage{amsmath}

\usepackage{epsfig}

\newcommand{\betac}{\beta_{\mathrm{c}}}
\newcommand{\betacdual}{\beta_{\mathrm{c}}^{\ast}}
\newcommand{\betaest}{\beta_{\mathrm{est}}}
\newcommand{\omegac}{\omega_{\mathrm{c}}}
\newcommand{\dm}{d_{\mathrm{m}}}
\newcommand{\Mrho}{M_{\rho}}
\newcommand{\arsinh}{\operatorname{arsinh}}
\newcommand{\IC}{the IC algorithm}

\newcommand{\scal}[1]{\langle #1 \rangle}
\newcommand{\arttit}[1]{#1}
\newcommand{\rem}[1]{}

\begin{document}

\jl{1}

\title[Invaded cluster algorithm for critical 
properties of planar Ising models]%
{Invaded cluster algorithm for critical properties of
periodic and aperiodic planar Ising models}

\author{Oliver Redner and Michael Baake}

\address{Institut f\"ur Theoretische Physik, Universit\"at T\"ubingen,
    Auf der Morgenstelle 14, D-72076 T\"ubingen, Germany}

\begin{abstract}
We demonstrate that the invaded cluster algorithm, recently introduced
by Machta \textit{et al}, is a fast and reliable tool for 
determining the critical temperature and the magnetic critical exponent of
periodic and aperiodic ferromagnetic Ising models in two dimensions.
The algorithm is shown to reproduce the known values of the critical
temperature on various periodic and quasiperiodic graphs with an
accuracy of more than three significant digits,
but only modest computational effort.
On two quasiperiodic graphs which were not investigated in this
respect before, the twelvefold symmetric square-triangle tiling and
the tenfold symmetric T\"ubingen triangle tiling, we determine the
critical temperature.
Furthermore, a generalization of the algorithm to non-identical
coupling strengths is presented and applied to a class of Ising
models on the Labyrinth tiling.
For generic cases in which the heuristic Harris-Luck criterion
predicts deviations from the Onsager universality class, we find a
magnetic critical exponent different from the Onsager value.
But also notable exceptions to the criterion are found which consist
not only of the exactly solvable cases, in agreement 
with a recent exact result, but also of the self-dual ones and maybe more.
\end{abstract}


\submitted


\section{Introduction}

Explicit exact solutions for the thermodynamic behaviour of (ferromagnetic)
Ising models are available, or expected to exist, only in one dimension
and for a few models in two dimensions.
For more complex cases, and in higher dimensions, one has to rely on
approximate techniques or simulations.
As one is usually interested in the expectation value of certain
observables such as the magnetization and the susceptibility at a
given temperature, Monte Carlo algorithms sampling the canonical
ensemble are often the numerical methods of choice.

The limiting factor in investigating the critical behaviour of spin systems
with algorithms of this kind is the critical slowing at the phase
transition point.  The time
needed to sample statistically independent system configurations
diverges with the system size according to a power law.  
The algorithms being most efficient in this respect are the so-called
cluster algorithms, the 
first of which was introduced by Swendsen and Wang in 1987
\cite{SwWa}.  
A single-cluster version, which shows less critical slowing in
three and more dimensions, was proposed by Wolff two years later \cite{Wol}.  
Also series expansions are suitable for studying the critical
behaviour of periodic models,  but they do not lead to satisfactory results
for aperiodic ones \cite{HighTemp}.

Estimating the critical temperature of the infinite system with Monte
Carlo algorithms is usually done using the Binder cumulant method
\cite{Bin}.  (E.g.\ this has been used in the first paper on the
universality class for quasiperiodic graphs \cite{Bha} which will play
an important role also in this paper.)  For a number of finite system
sizes and several temperature values taken from some interval
containing the critical temperature, the fourth cumulant of the
magnetization is measured.  
This cumulant is zero for infinite temperature, $\frac{2}{3}$ for zero
temperature and depends on the system size for all other temperature
values; only at the critical temperature the curves for different
system sizes intersect. 
This admits rather reliable determination of the location of the
critical point.  However, because
the system has to be simulated at several temperature values, this
procedure is very time-consuming, and a rough knowledge of the
location of the critical temperature is required beforehand.
So, independent methods are certainly useful.

In 1995 Machta \etal \cite{IC1} developed a self-organized version of the
Swendsen-Wang algorithm
which they dubbed \emph{invaded cluster} (IC) algorithm.  
It is not only able to locate the critical temperature without prior
knowledge and without any temperature sweeps being necessary, but also
seems to show yet less critical slowing than the Wolff algorithm
\cite{ICdyn}.
The algorithm was shown to reproduce three significant digits of the
known values of the critical temperature on the square and simple
cubic lattice Ising models with modest computational effort
\cite{IC1,IC2}.
The ensemble that is sampled, however, is not the canonical one, and not
much is known about it rigorously.  In 
the thermodynamic limit, it is expected to be equivalent to the latter,
but about the finite-size scaling \cite{Bar} nothing helpful is known.
Also, the thermal exponent and equivalent quantities,
such as $\nu$ and $\alpha$, are not easily measured with \IC\ 
\cite{ICdyn,Fra}.

This lack of theoretical knowledge is responsible for the fact that 
so far the best accuracy in measuring the critical temperature is
still an order of magnitude lower 
than with other cluster algorithms \cite{ICpar}.  
Nevertheless, at this accuracy, \IC\ is substantially faster.  This is
mainly due to the fact that the temperature need not be varied.  In
addition, the autocorrelation times are by a factor of 4 or more
smaller than those of the Wolff algorithm \cite{IC2}.  (If one is also
interested in observables on sub-systems this is not true, but the
autocorrelation times are still
comparable to the ones of the Wolff algorithm, see \cite{ICdyn}.)
As a rule of thumb, one can expect \IC\ to be at least a factor
of 5 faster.   
Therefore it is a good choice for obtaining
a good estimate of the critical temperature with modest computational
effort.  In addition, one of the two independent critical exponents,
the magnetic exponent $y_h$, can be measured easily.

Those quantities are of particular interest
for Ising models on quasiperiodic tilings, a class of graphs being
used as models for the structure of quasicrystals, compare
\cite{MathQC,Aper} and references therein.
As these graphs are very homogeneous, the influence of the
quasiperiodicity on the critical temperature is expected to be small
and its rough location should be governed by the first mean coordination
number, see \cite{Coord} and references therein. 
The critical exponents are expected to be the same as 
their periodic counterparts unless the degree of disorder is bigger
than a `critical' value.  
This is the result of the heuristic criterion by Luck \cite{Luc} who
generalized the well-known Harris criterion for random disorder
\cite{Har} to aperiodic structures.
This criterion, now called the Harris-Luck criterion, was recently
proved for certain kinds of one-dimensional disorder  
\cite{HerQC,HerXY} and corroborated by approximate methods for 
two-dimensional substitution systems \cite{HerDiss,HerICQ}.  But in two
or more dimensions, the validity has not yet been systematically
confirmed by simulations. 

The IC algorithm has so far only been tested on the square and the simple cubic
lattice -- both with success \cite{IC1,IC2,ICpar,ICdyn}.  Before its
results for more complex 
situations can be trusted, however, it should be tested on other
graphs for which the critical temperature is known.  
This will be done in \sref{sec:test} for a number of periodic and
quasiperiodic graphs.
Then, as a first application, the critical temperature of the Ising model on
two quasiperiodic tilings, the twelvefold symmetric square-triangle tiling
\cite{STT} and the tenfold symmetric T\"ubingen triangle tiling
\cite{TTT}, will be determined.
In \sref{sec:gener}, we will present a generalization of \IC\ to
models with arbitrary coupling strengths.
Results of simulations on different realizations of the Labyrinth
tiling \cite{Sir} with this generalized algorithm will be described in
\sref{sec:laby}. 
\Sref{sec:summary} will give a summary and discussion.


\section{The critical temperature of models with identical couplings}
\label{sec:test}

We consider the field-free Ising model defined by a spin
$\sigma_i=\pm1$ on each vertex $i$ of some graph and a ferromagnetic bond
$\scal{i,j}$ of coupling strength $J_{ij}>0$ between each
pair of neighbouring spins.
The internal energy is
$\mathcal{H}(\bsigma)=-\sum_{\scal{i,j}} J_{ij}\sigma_i\sigma_j$ 
where the sum is over all bonds.
In this section, we restrict ourselves to identical couplings
$J_{ij}\equiv1$.

For the Ising model with this restriction, the critical temperature is
known exactly on a number of periodic graphs, see \cite{Syo} for a survey.  
In the case of two quasiperiodic tilings, the Penrose tiling
\cite{Pen} and the octagonal Ammann-Beenker tiling \cite{Amm,Bee},
there exist Monte Carlo estimates \cite{Oka,Soe,Led} and
high-precision numerical values from a recent
analysis of large periodic approximants using Kac-Ward determinants
\cite{KacWard}.  
With those values, also the critical temperature on the corresponding
dual tilings is known through the exact relationship \cite{Syo} 
\begin{equation}
\label{eq:duality}
\sinh(2\betac)\sinh(2\betacdual)=1,
\end{equation}
where $\betac$ and $\betacdual$
are the inverse critical temperatures on a graph and its dual.
All this offers a number of graphs on which \IC\ can be
tested thoroughly.

\begin{figure}
\begin{indented}
\item[]%
\input{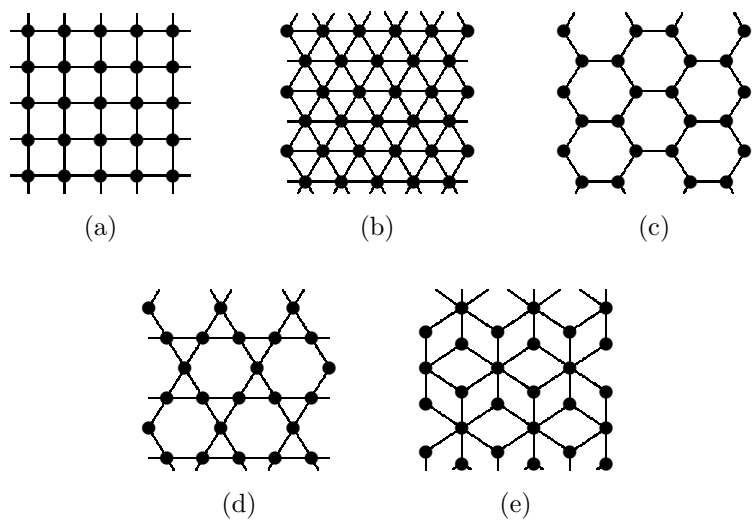}
\end{indented}
\caption{Parts of the periodic graphs considered here:
(a) square lattice, (b) triangular lattice, 
(c) hexagon packing, (d) Kagom\'e graph, and
(e) diced graph.}
\label{fig:periodic}
\end{figure}

\begin{figure}
\begin{indented}
\item[]%
\begin{minipage}[b]{140pt}
  \centering\epsfig{file=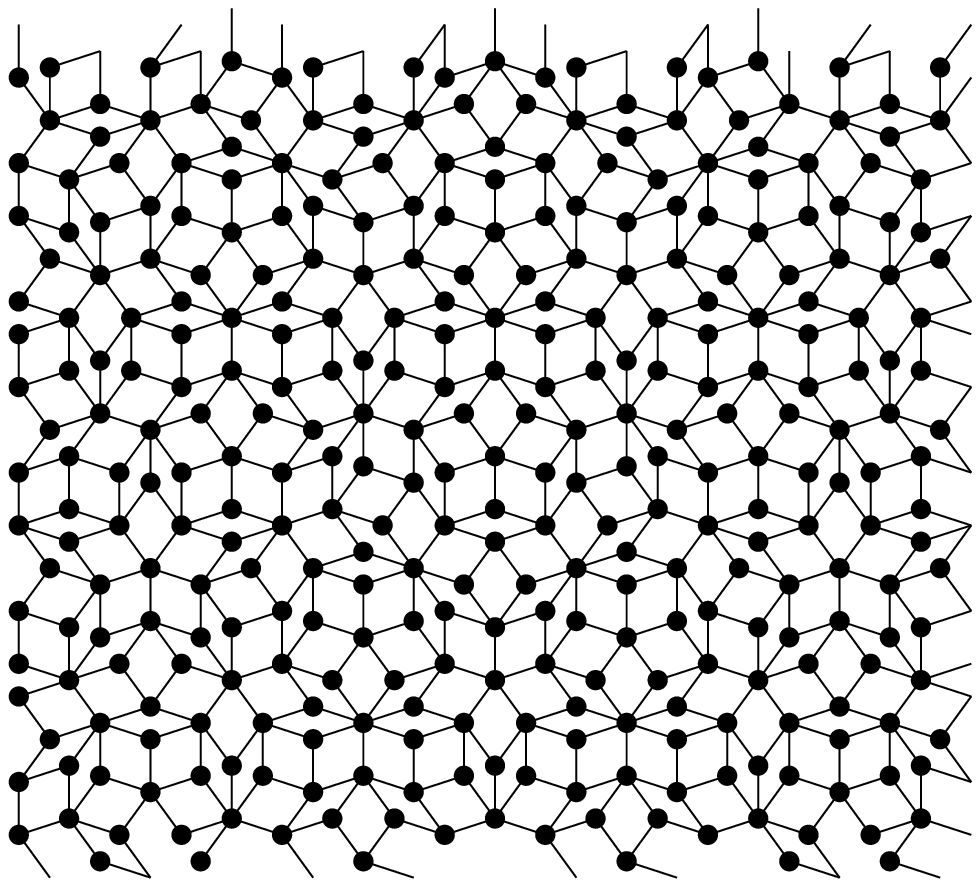,width=140pt,clip=,%
  bbllx=16,bblly=33,bburx=176,bbury=193}
  (a)
\end{minipage}\hspace*{4ex}
\begin{minipage}[b]{140pt}
  \centering\epsfig{file=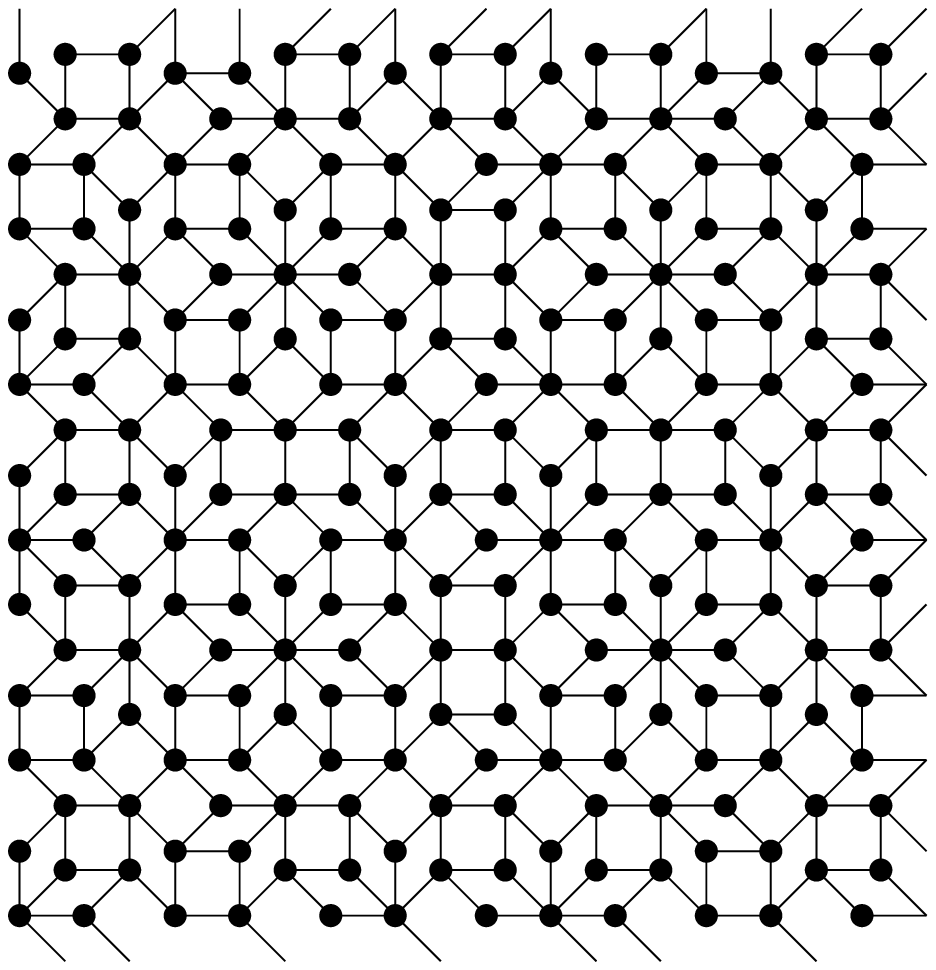,width=140pt,clip=,%
  bbllx=20,bblly=28,bburx=210,bbury=218}
  (b)
\end{minipage}\\[2ex]
\begin{minipage}[b]{140pt}
  \centering\epsfig{file=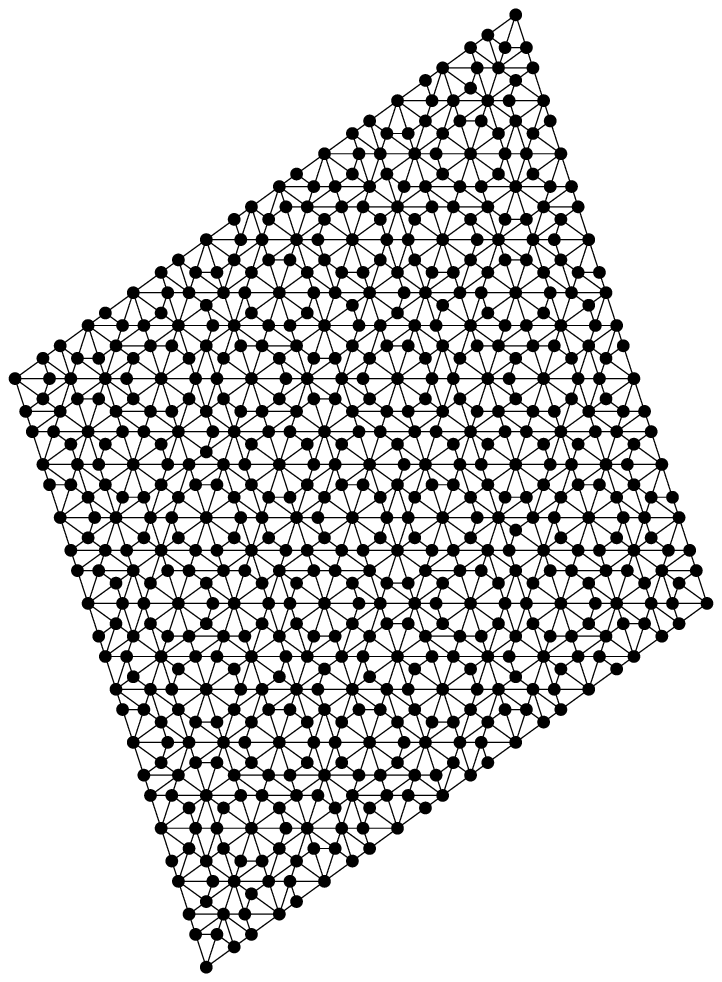,width=140pt,clip=,%
  bbllx=84,bblly=84,bburx=184,bbury=184}
  (c)
\end{minipage}\hspace*{4ex}
\begin{minipage}[b]{140pt}
  \centering\epsfig{file=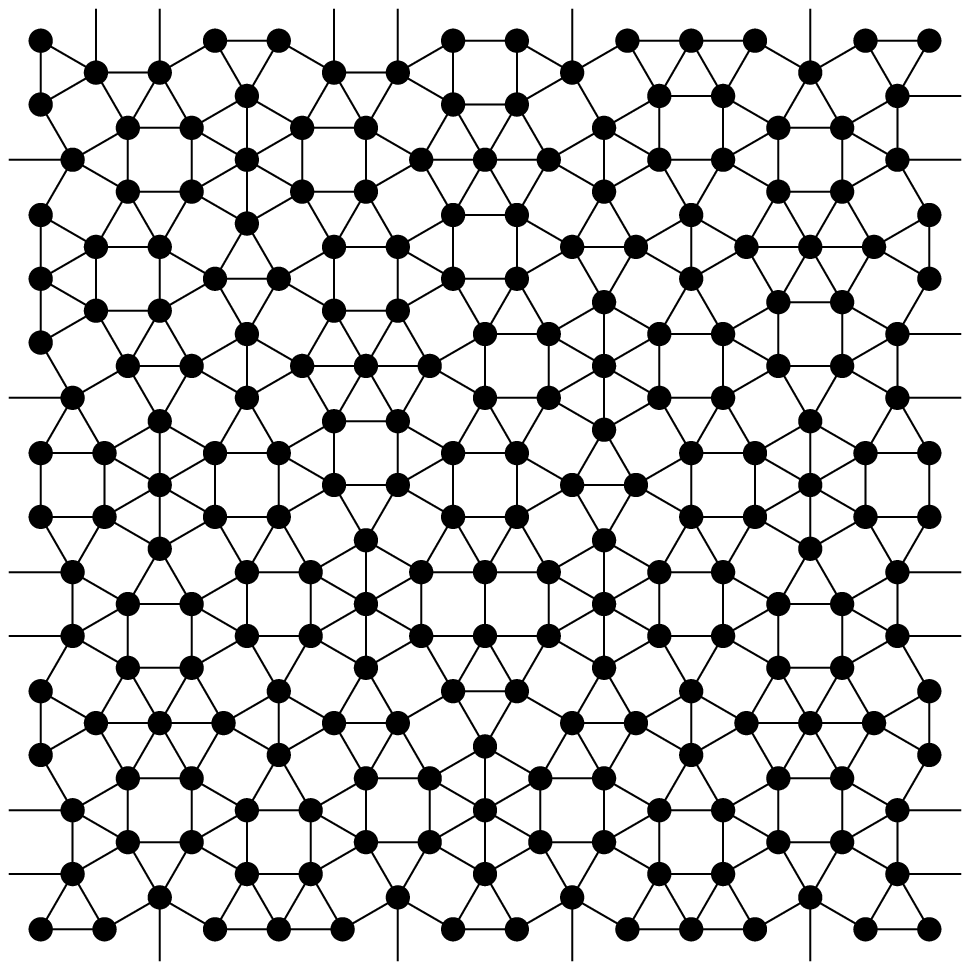,width=140pt,clip=,%
  bbllx=30,bblly=24,bburx=220,bbury=214}
  (d)
\end{minipage}
\end{indented}
\caption{Parts of the quasiperiodic tilings considered here:
(a) Penrose, (b) Ammann-Beenker,
(c) T\"ubingen triangle, and
(d) twelvefold square-triangle tiling.}
\label{fig:tilings}
\end{figure}

With \IC, a given configuration of a (finite-sized) system is updated
in two steps.  First, 
bonds between aligned spins (\emph{satisfied} bonds) are selected in
random order until one connected component (\emph{cluster}) wraps
around the system in at least one dimension (periodic boundary
conditions are assumed) or all satisfied bonds have been selected.   
Then, each cluster (including each isolated spin) is flipped with
probability $\frac{1}{2}$. 
The fraction of selected to satisfied bonds, $f$, gives an estimate
$\betaest$ of the inverse critical temperature via \cite{IC1}
\begin{equation}
\label{eq:f}
f=1-e^{-2\betaest}.
\end{equation}
This is reasonable as the expectation value of $f$ in the canonical
ensemble is just the probability $p=1-e^{-2\beta}$ to select a bond
in the Swendsen-Wang algorithm.

For each periodic graph depicted in \fref{fig:periodic}, we used \IC\
to simulate the Ising model at 13 different 
linear system sizes $L$ between 16 and 288.  For the quasiperiodic
tilings, shown in \fref{fig:tilings}, we had to restrict ourselves to
the periodic approximants available. 
With each size, 10 bunches of 5,000 update steps each were taken.
This took approximately $2.3\cdot10^{-6}$ seconds per bond 
on a 266 MHz Pentium II processor which amounts to roughly one day
for all sizes of one graph. 
The first bunch was used for equilibration and then discarded.  For each
one of the remaining, the averages of the quantities of interest were
computed and those numbers treated as an independent sample.

\begin{figure}
\begin{indented}
\item[]%
\epsfig{file=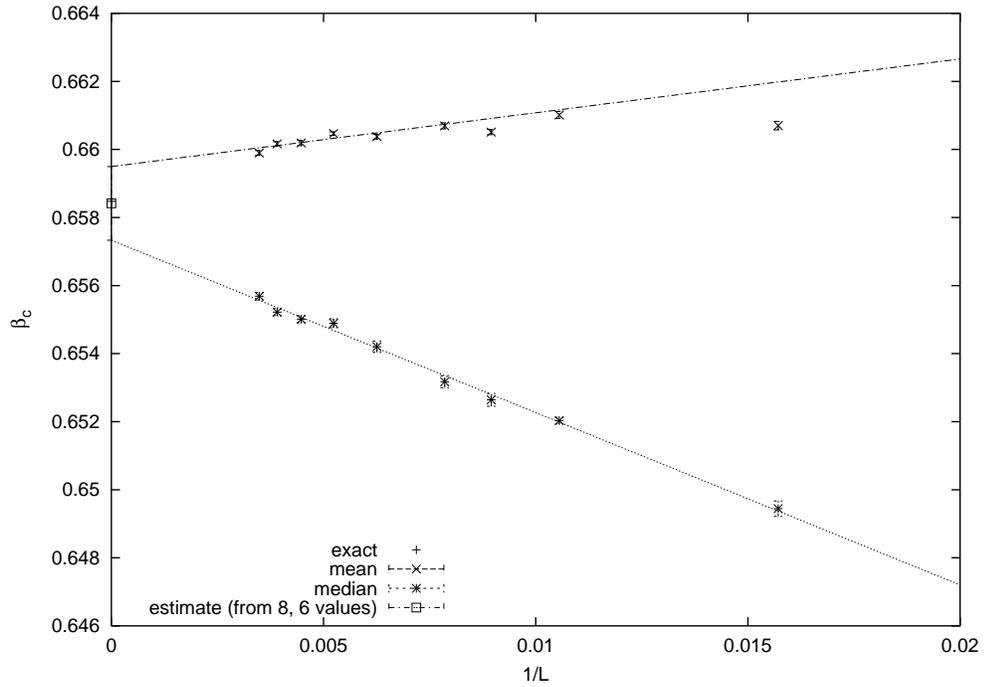,angle=-90,width=375pt}
\end{indented}
\caption{Finite-size scaling of the IC estimates of the inverse critical
  temperature on the hexagon packing.}
\label{fig:fschexa}
\end{figure}

In contrast to the previous IC studies \cite{ICpar,IC1,IC2,ICdyn}, in
which the average of $f$ was determined and $\betaest$ inferred in the
end using \eref{eq:f}, we computed $\betaest$ in each step.  This
leads to the same results after finite-size scaling, as shown in
\cite{Red}.  \Fref{fig:fschexa} shows the results for the estimates of
the inverse critical temperature on the hexagon packing as a typical
example.  The picture is qualitatively the same for all graphs
considered, details can be found in \cite{Red}.  Due to certain
bottlenecks occuring in the cluster growth process \cite{IC2}, the
mean value of $\betaest$ is always greater than the median.  But both
are expected to approach the same value $\betac$ in the thermodynamic
limit $L\rightarrow\infty$.  From the data it is obvious that both
curves are inclined towards each other.  This gives rise to the
following recipe for estimating $\betac$ and its error.  We compute
linear fits to both the mean and median values independently using the
data points corresponding to some of the largest system sizes.  Then
we take the mean of both $\betac$-axis intercepts as an estimate for
$\betac$ and half their separation as its error.

There is no a priori justification for this recipe besides the
numerical evidence that $\betac$ lies in between both points.
The results shown in \tref{tab:betac} agree with the known values
to within more than three significant digits.
This indicates that the way $\betac$ is estimated is reasonable and
that we considerably overestimate the error. 
For the Penrose and the Ammann-Beenker tiling, the agreement of our
data with the high-precision values \cite{KacWard} is even better than the
one of the other Monte 
Carlo values \cite{Oka,Soe,Led}.  This is especially remarkable if
one takes into account the small computational effort.

\begin{table}
\caption{IC estimates of the inverse critical temperature for
  various graphs.  Note that the errors given are very
  conservative upper bounds.  Exact and other numerical values are given for
  comparison where available.
  The exact values are taken from \cite{Syo}, the best numerical ones from
  \cite{KacWard}.  The values for the dual tilings marked with *
  were inferred using the exact relation \eref{eq:duality}.}
\label{tab:betac}
\lineup
\footnotesize\rm
\begin{tabular}{@{}lllll}
\br
graph & \centre{3}{$\betac$} & (mean)\\
 & \crule{3} & coordination\\
 & exact/best & IC estimate & other values & number\\
\mr
square lattice & $0.4406868\ldots$ & 0.4405(4) & & 4\\
triangular lattice & $0.2746531\ldots$ & 0.2747(5) & & 6\\
hexagon packing & $0.6584789\ldots$ & 0.6584(11) & & 3\\
Kagom\'e graph & $0.4665660\ldots$ & 0.4665(8) & & 4\\
diced graph & $0.4157215\ldots$ & 0.4157(7) & & 4\\
\ms
Penrose tiling & 0.417046(1) \rem{KacWard} & 0.4170(8) & 0.4181(7)\0
 \cite{Oka} & 4\\
 & & & 0.4165(9)\0 \cite{Soe} \\
dual Penrose tiling & 0.465145(1)* \rem{KacWard} & 0.4652(8) & 
  & 4\\
Ammann-Beenker tiling & 0.41887800(1) \rem{KacWard} & 0.4191(7) &
  0.4186(7)\0 \cite{Led} & 4\\
dual Ammann-Beenker & 0.46318974(1)* \rem{KacWard} & 0.4634(10) &
  & 4\\
\ms
T\"ubingen triangle tiling & & 0.2565(6) & & 6\\
square-triangle tiling & & 0.3364(5) & & 
$5.072\ldots\simeq12-4\sqrt{3}$ \cite{Bri}\\
\br
\end{tabular}
\end{table}

One can conclude from this that the values of the critical temperature for
the twelvefold symmetric square-triangle tiling \cite{STT} and the
tenfold symmetric T\"ubingen triangle tiling \cite{TTT} given in
\tref{tab:betac} should have a similar accuracy.
Also, they are in the correct rough range that one expects from the
comparison of their first mean coordination number with the one of the
square and triangular lattice, see \tref{tab:betac}.

The autocorrelation times found were all compatible with the ones
given in table II of \cite{ICdyn}, especially they were always less
than unity.  Therefore it seems safe to conclude that \IC\ is not
significantly slower on quasiperiodic graphs than on periodic ones.
In particular, it should still be much faster in finding the critical
temperature (to the above mentioned accuracy) than other Monte Carlo
algorithms.


\section{Generalization of \IC\ to non-identical coupling strengths}
\label{sec:gener}

The IC algorithm was so far only published for identical coupling strengths
$J_{ij}\equiv J$, but a generalization to arbitrary ones is rather
straightforward \cite{Mac}.
To this end, let us review how the Swendsen-Wang (SW) and IC algorithms
are connected.
In the SW algorithm, for each satisfied bond $\scal{i,j}$ a random
number $0\le t_{ij}<1$ is drawn from a uniform distribution and the
bond is selected if the condition
\begin{equation}
\label{eq:SW}
t_{ij} < p_{ij} := 1 - e^{-2\beta J_{ij}}
\end{equation}
is fulfilled.  In \IC, on the contrary, bonds are selected in random
order until one cluster wraps around the system (or a different
condition is met, see \cite{IC2,ICpar}).  After solving \eref{eq:SW}
for $\beta$,
\begin{equation}
\label{eq:beta}
\beta < -\frac{1}{2 J_{ij}}\log(1-t_{ij}) =: \beta_{ij},
\end{equation}
we see how the algorithm has to be generalized.
We again draw a random number $t_{ij}$ for each bond and calculate the
corresponding $\beta_{ij}$ from \eref{eq:beta}.
Then we sort the bonds ascendingly with regard to $\beta_{ij}$ and
select the satisfied ones in this order until one cluster wraps around
the system (or all satisfied bonds have been selected).  If bond
$\scal{k,l}$ was selected last, $\beta_{kl}$ 
gives an estimate $\betaest$ for the inverse critical temperature.

The random numbers $t_{ij}$ play a twofold role here.  They determine
the order in which the bonds are selected, and the one corresponding to
the last bond selected gives the numerical value of $\betaest$.
So it is not necessary to use independent distributions for
the $t_{ij}$ as long as the marginal distributions are the same for
all of them.
Instead, one can use a random permutation of equidistant points
`approximating' the interval $[0,1[$ for the $t_{ij}$.  
With such a choice, the original algorithm is recovered for identical
couplings (in which case the sorting is trivial).

Now, there is a crucial difference between the original and the
generalized version with regard to the computational complexity.  In
the case of identical couplings, only a permutation of the $N$ bonds has to
be created.  This can be done with a computational effort of order
$\Or(N)$.  The sorting of the arbitrary numbers $\beta_{ij}$,
in contrast, requires an effort of $\Or(N\log N)$ which makes the algorithm
considerably slower.
If, however, only a small (constant) number $k$ of different coupling strengths
is present, the sorting can be done with effort $\Or(kN)=\Or(N)$ by
the following steps.
(Let us assume all bonds to be numbered sequentially $1,\dots,N$ here.)
\begin{enumerate}
\renewcommand{\labelenumi}{\arabic{enumi}.}
\item  Create an auxiliary permutation $\pi$ of the numbers $\{1,\dots,N\}$.
\item  For each bond, define a value of the inverse temperature,
  \[
  \beta_{\pi(i)} := 
    -\frac{1}{2 J_{\pi(i)}}\log\left(1-\frac{2i-1}{2N}\right),
  \quad i=1,\dots,N.
  \]
\item  Create the final permutation $\Pi$ of $\{1,\dots,N\}$
  by repeating the following steps for $i=1,\dots,N$ after
  initializing a list $(k_t)_{1\le t\le k}$ with the indices of the
  first bonds of type $t$ appearing in $\pi$.
  \begin{enumerate}
  \item  Determine the value of $t$ for which $\beta_{\pi(k_t)}$
    becomes minimal.
  \item  Set $\Pi(i):=\pi(k_t)$.
  \item  Replace $k_t$ by the next bond of type $t$ appearing in $\pi$.
  \end{enumerate}
\end{enumerate}
Steps 1 and 2 are possible with effort $\Or(N)$.  In step 3, each of
the $k$ counters $k_t$ takes on all the values $\{1,\dots,N\}$ in the
worst case.  This step can therefore also be performed with 
effort $\Or(kN)=\Or(N)$.  Overall, the sorting procedure takes linear
computational effort, only with a higher constant than in the
original algorithm. 

The generalized IC algorithm will be applied to a class of Ising
models with non-identical coupling strengths in the following section.


\section{The Labyrinth tiling as an example to study the relevance of disorder}
\label{sec:laby}

\subsection{The Ising model on the Labyrinth tiling}

\begin{figure}
\begin{indented}
\item[]%
\epsfig{file=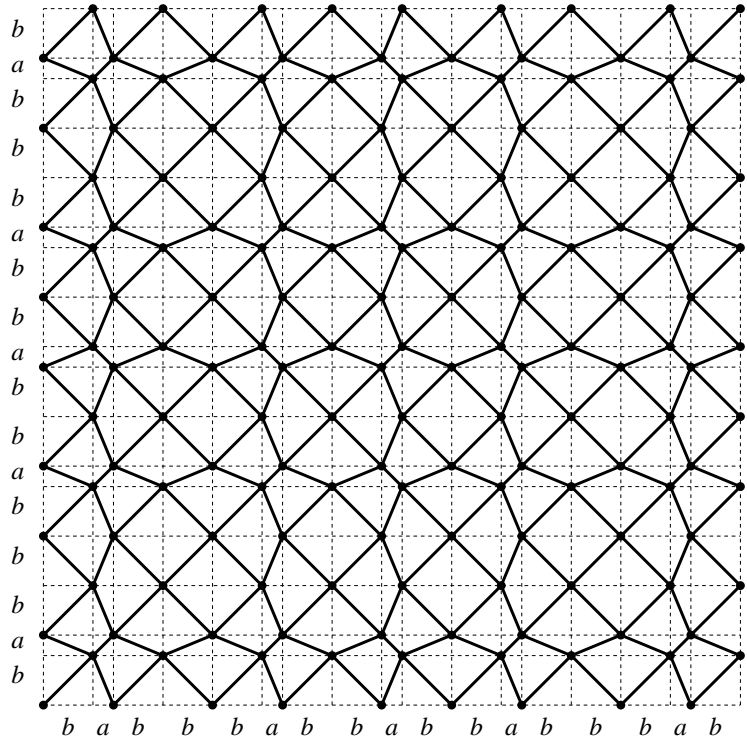,width=280pt}
\end{indented}
\caption{Part of the Labyrinth tiling corresponding to the
  substitution with $k=1$.}
\label{fig:laby}
\end{figure}

The Labyrinth tiling \cite{Sir} is constructed from a
substitution $\rho$ on the two-letter alphabet $\mathcal{A}=\{a,b\}$.  Here
we will consider substitutions of the type
\[
\rho:   \begin{array}{lll}
        a & \rightarrow & b \\ b & \rightarrow & ba^kb
        \end{array},
\]
with $k\ge1$.  The substitution is iterated infinitely, starting from
the word $b$ (or 
any other finite word), i.e.\ in each step it is applied to each
letter of the word.  This 
yields a (semi\discretionary{-)}{}{-)}infinite limiting word $w$.  
For $k=1$, we have the silver mean chain treated in \cite{BGB}.  
Each letter is assigned an interval of a specific length.  
In this representation, the Cartesian product of $w$ with
itself is taken giving one quadrant of a (distorted) square lattice.  
Of this, one takes every other vertex, starting from the origin, and
connects each one with its nearest neighbours over the diagonals of the
underlying cells. 
As there are four types of those cells according to the
letters $x,y\in\mathcal{A}$ on the vertical and horizontal copy of
$w$, and each bond can 
be either `raising' or `lowering', we can distinguish 8 different types
of bonds in the corresponding Ising model.  We will denote the
corresponding coupling strengths by $J_{xy}$ for the raising and
$\tilde{J}_{xy}$ for the lowering bonds.  Part of the Labyrinth
tiling corresponding to $k=1$ is shown in \fref{fig:laby}.

On a 4-dimensional submanifold of the 8-dimensional coupling space, the
Ising model is exactly solvable and shows critical behaviour of
Onsager type regardless of the underlying word $w$ \cite{BGB}.  
To see what the Harris-Luck criterion \cite{Har,Luc} predicts, we have
to look at the fluctuation exponent $\omega$, which describes how the
deviations of the mean coupling strength in a finite patch, compared to the
infinite volume mean, scale with the size of the patch.  If $\omega$ is
greater than the `critical' value 
$\omegac=1-\frac{1}{\dm\nu}$ \cite{Luc}, one expects the disorder to be
\emph{relevant}, i.e.\ critical exponents deviating 
from the Onsager values.  Here, $\nu$ is the correlation length
exponent of the periodic system ($\nu=1$ in our case) and the disorder
affects all $\dm=2$ dimensions, thus $\omegac=\frac{1}{2}$.
For $\omega<\omegac$, the disorder is expected to be
\emph{irrelevant}.  In the case of \emph{marginal} disorder,
$\omega=\omegac$, the criterion does not make any specific predictions.

For substitution systems, the fluctuation exponent $\omega$ can be extracted
from the bond substitution, which is in our case induced by the letter
substitution $\rho$.  It is given by $\omega = \log|\lambda_2|/\log\lambda_1$,
where $\lambda_1>|\lambda_2|$ are the two largest eigenvalues
of the
corresponding substitution matrix whose entries count how many bonds of one
type (determined by the row of the matrix) are in the substitute for
another type (the column).
This matrix turns out to be described by the tensor product of the substitution
matrix 
$\Mrho=\left(\begin{smallmatrix} 0 & k \\ 1 & 2 \end{smallmatrix}\right)$
for $\rho$ with itself, its spectrum being 
$\sigma(\Mrho\otimes\Mrho)=\{\lambda\mu: \lambda,\mu\in\sigma(\Mrho)\}$
with $\sigma(\Mrho)=\{1\pm\sqrt{1+k}\}$.
Thus, we find the fluctuation exponent
\begin{equation}
\label{eq:omega}
\omega = \frac{\log k}{2\log(1+\sqrt{1+k})}.
\end{equation}
Substitutions with $k<3$ correspond to irrelevant disorder, $k=3$ is
the marginal case, and for $k>3$ the disorder should be relevant.
These predictions shall be tested shortly.

As an alternative model, let us also look at a random case.
Instead of using a substitution for creating the word $w$, one can
also use a random word of which each letter is independently chosen to be
$a$ or $b$ with probability $p_a$ and $(1-p_a)$, respectively.  Due to
the way the Labyrinth is constructed, the resulting bond distribution 
has correlations within each row and column.
Accordingly, this kind of disorder has a higher fluctuation exponent
than uncorrelated disorder, $\omega=\frac{3}{4}$
compared to $\omega=\frac{1}{2}$, and is relevant with respect to the
Harris-Luck criterion.


\subsection{Results}

First, we successfully tested the implementation of the generalized
algorithm on graphs with identical couplings and a few of the exactly
solvable cases of the Labyrinth tiling.

\begin{table}
\caption{Choices for the coupling strengths used in the simulations for
  the Labyrinth tiling.  The ones for the raising and lowering bonds are
  denoted by $J_{xy}$ and $\tilde{J}_{xy}$,
  respectively, where $x$ and $y$ are the corresponding letters in the
  horizontal and vertical copy of the word $w$.}
\label{tab:choices}
\setlength\tabcolsep{5.3pt}
\footnotesize\rm
\begin{tabular}{@{}lllllllll@{}}
\br
choice & $J_{aa}$ & $J_{ab}$ & $J_{ba}$ & $J_{bb}$ & $\tilde{J}_{aa}$
 & $\tilde{J}_{ab}$ & $\tilde{J}_{ba}$ & $\tilde{J}_{bb}$ \\
\mr
\#1 & 1.61268    & 1         & 0.63693035 & 0.35675743 
    & 0.55876724 & 1         & 1.4707784  & 2.1084272  \\
\#2 & 1.3403778  & 1         & 0.77308147 & 0.55051591 
    & 0.71949498 & 1         & 1.2644651  & 1.6289055  \\
\#3 & 0.4        & 0.8       & 1          & 2.8
    & 1.9809928  & 1.2285752 & 1          & 0.19281531 \\
\#4 & 2.5        & 1         & 0.5        & 0.75
    & 0.25159591 & 1         & 1.7343053  & 1.2964035  \\
\#5 & 1.2        & 1         & 0.7        & 2
    & 1.4        & 0.9       & 1.1        & 1.8        \\
\#6 & 2.5        & 1         & 0.4        & 0.8
    & 2.8        & 0.9       & 1.3        & 0.7        \\
\#7 & 0.4        & 0.8       & 1          & 2.8
    & 1.3        & 2.3       & 0.9        & 2.4        \\
\#8 & 2          & 1.2       & 1.2        & 0.5
    & 2          & 1.2       & 1.2        & 0.5        \\
\br
\end{tabular}
\end{table}

Then, we applied \IC\ to the Ising model 
on the Labyrinth corresponding to the substitutions $k=2$, 3, 4 and
on the random version with $p_a=0.4$, using 
the choices for the coupling strengths given in \tref{tab:choices}.
The first two choices (\#1, \#2) are
on the exactly solvable submanifold, i.e.\ we chose three of the
coupling strengths ($J_{aa}$, $J_{ab}$, $J_{ba}$) arbitrarily and
the inverse critical
temperature to be the one of the square lattice Ising model,
$\betac=\arsinh(1)/2\simeq0.44069$, and determined the
other five couplings by numerically solving the five equations
\cite{BGB}
\begin{equation}
\label{eq:solv1}
\sinh(2\betac J_{xy})\sinh(2\betac\tilde{J}_{xy})=1,\quad x,y\in\mathcal{A},
\end{equation}
and
\begin{eqnarray}
\label{eq:solv2}
\left(\frac{\sinh(2\betac J_{aa})\sinh(2\betac J_{bb})}%
{\sinh(2\betac J_{ab})\sinh(2\betac J_{ba})}\right)^2
\frac{\cosh(\betac[-\tilde{J}_{aa}+J_{ba}+\tilde{J}_{bb}+J_{ab}])}%
{\cosh(\betac[-J_{aa}+\tilde{J}_{ba}+J_{bb}+\tilde{J}_{ab}])}\times\nonumber\\
\quad\times\frac{\cosh(\betac[J_{aa}-\tilde{J}_{ba}+J_{bb}+\tilde{J}_{ab}])
\cosh(\betac[\tilde{J}_{aa}+J_{ba}-\tilde{J}_{bb}+J_{ab}])}%
{\cosh(\betac[\tilde{J}_{aa}-J_{ba}+\tilde{J}_{bb}+J_{ab}])
\cosh(\betac[J_{aa}+\tilde{J}_{ba}-J_{bb}+\tilde{J}_{ab}])}\times\\
\quad\times\frac{\cosh(\betac[J_{aa}+\tilde{J}_{ba}+J_{bb}-\tilde{J}_{ab}])}%
{\cosh(\betac[\tilde{J}_{aa}+J_{ba}+\tilde{J}_{bb}-J_{ab}])}=1.\nonumber
\end{eqnarray}
The first four \eref{eq:solv1} are just the duality conditions on the
lowering and raising bonds of the same type.
The fifth one \eref{eq:solv2} involves all the remaining four coupling
strengths. 
The next two choices (\#3, \#4) are self-dual but not exactly
solvable, i.e.\ they fulfill the duality
conditions \eref{eq:solv1} with respect to our chosen temperature, but
the fifth condition \eref{eq:solv2} is significantly violated. 
The last four choices (\#5 to \#8) violate all five conditions. 
Choice \#8 is isotropic, i.e.\ bonds are only distinguished
with regard to their length.

\begin{table}
\caption{IC estimates of the magnetic critical exponent $y_h$
  on the Labyrinth tiling for three different substitutions and the
  random case, each one with the eight choices of the coupling
  strengths given in \tref{tab:choices}.
  The value marked with * is significantly different from the Onsager
  value $y_h=1.875$.  Due to fluctuations as visible in part (b) of
  \fref{fig:mass}, the total error for the values marked with * or \dag\
  is presumably quite large ($\approx0.03$).  In the random case, the
  flucuations for choices \#5 to \#8 were too large to even confirm a
  power-law behaviour.
  The fluctuation exponent $\omega$ for the substitutions
  was computed using \eref{eq:omega}.}
\label{tab:yh}
\begin{indented}
\item[]%
\setlength\tabcolsep{5.9pt}%
\begin{tabular}{@{}llllllllll@{}}
\br
    &          & \centre{8}{choice of couplings}\\
$k$ & $\omega$ & \#1 & \#2 & \#3 & \#4 & \#5 & \#6 & \#7 & \#8 \\
\mr
2 & $0.345\ldots$ 
    & 1.872 & 1.871 & 1.875 & 1.876 & 1.875 & 1.871     & 1.859  & 1.872\\
3 & $\frac{1}{2}$ 
    & 1.874 & 1.875 & 1.873 & 1.876 & 1.871 & 1.847\dag & 1.732\dag & 1.880\\
4 & $0.590\ldots$ 
    & 1.872 & 1.873 & 1.877 & 1.875 & 1.866 & 1.839\dag & 1.453* & 1.898\dag\\
random & $\frac{3}{4}$ 
    & 1.874 & 1.878 & 1.877 & 1.862 
    & \centre{4}{\emph{fluctuations too large}}\\
\br
\end{tabular}
\end{indented}
\end{table}

\begin{figure}
\begin{indented}
\item[]%
\begin{minipage}[b]{375pt}
  \centering\epsfig{file=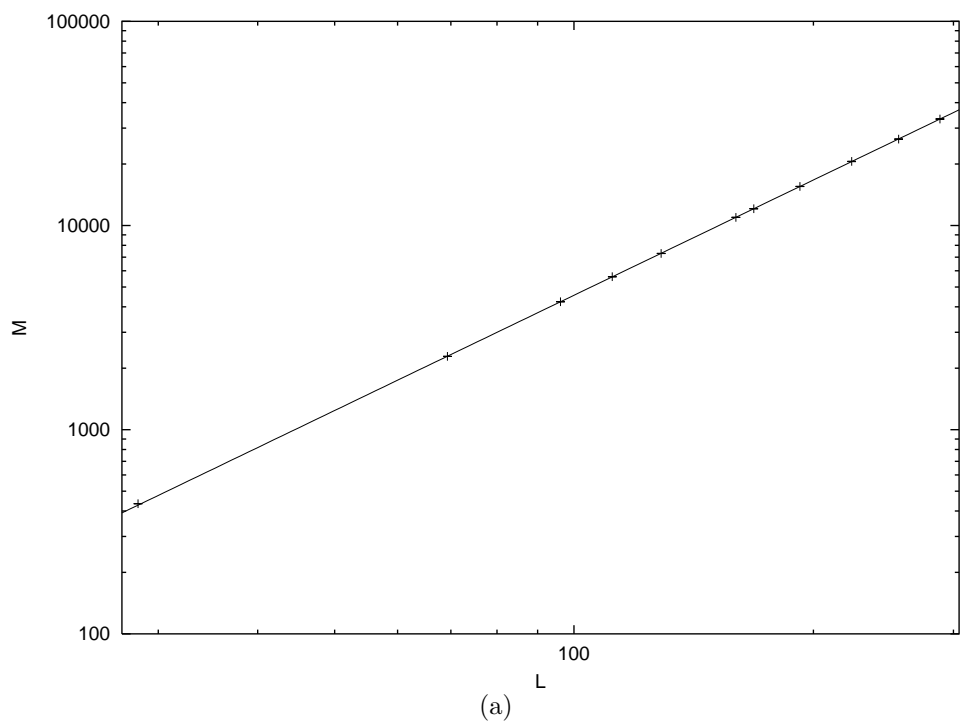,angle=-90,width=375pt}
  (a)
\end{minipage}\\[2ex]
\begin{minipage}[b]{375pt}
  \centering\epsfig{file=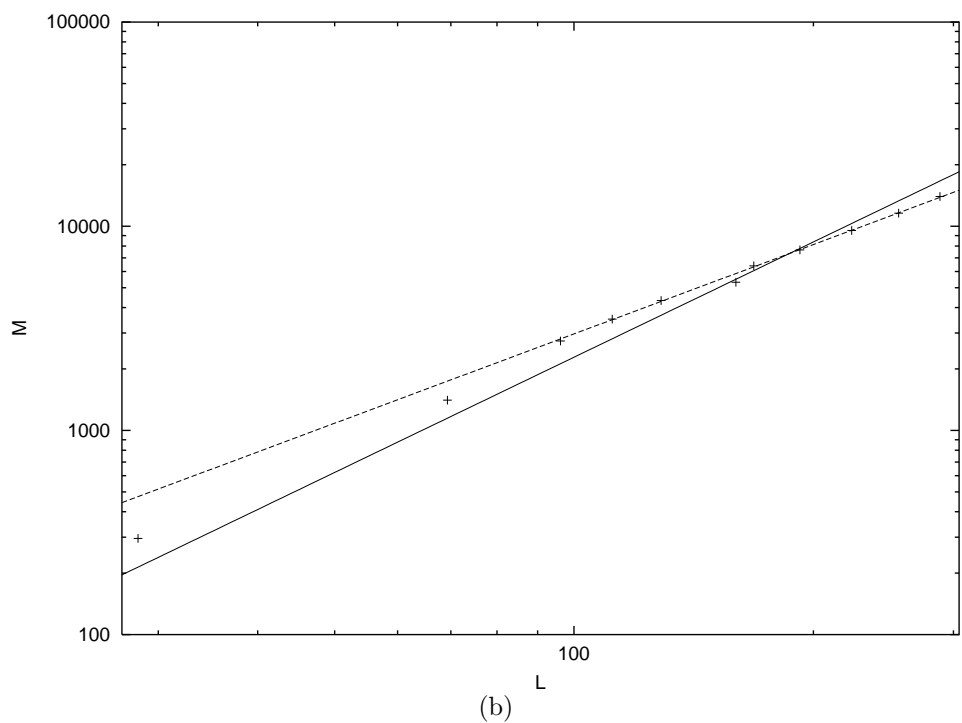,angle=-90,width=375pt}
  (b)
\end{minipage}
\end{indented}
\caption{Determination of the magnetic exponent $y_h$ on the Labyrinth
  tiling.  The mass $M$ of the cluster wrapping around the system is
  plotted versus the linear system size $L$. 
  (a) is a typical example of Onsager behaviour (coupling choice \#4
  with $k=4$) and (b) is typical of non-Onsager behaviour
  (choice \#7 with $k=4$).  The solid lines have the Onsager
  slope $y_h=1.875$, the dashed line is a linear fit yielding
  $y_h\simeq1.453$.
  The fluctuations in (b) are due to variations in the bond
  frequencies in the approximants used.}
\label{fig:mass}
\end{figure}

We estimated the magnetic critical exponent $y_h=2-\beta/\nu$
as described in \cite{ICpar}
by plotting the logarithm of the expectation value of
the mass $M$ of the cluster that wrapped around the system against the
logarithm of the linear system size $L$ and then taking a linear fit. 
As the expectation values of $M$ and the magnetization of the system
are the same, the usual scaling relation for the magnetization becomes
$M\propto L^{y_h}$.
Two examples are shown in \fref{fig:mass}, the results are summarized
in \tref{tab:yh}. 
For the exactly solvable choices (\#1, \#2), we always found 
the exponent to be of Onsager type and the critical temperature 
to agree with the exact solution, i.e.\ with the choice made above.
For the self-dual choices (\#3, \#4), again, the Onsager exponent was
found in all cases.  The critical 
temperature, however, was not the chosen one for even values of $k$
(2, 4), but agreed with it for odd ones (1, 3, 5) and probably also in
the random version.  (For the latter, the fluctuations in $\betaest$
were too large to state this unambiguously.)  This is not yet understood.  

For the arbitrary choices (\#5 to \#8),
the critical exponent was of the Onsager class for $k=2$.  
For $k\ge3$, we still found power-law behaviour for all choices, but
a significantly different exponent for choice \#7 with $k=4$.
For the values in \tref{tab:yh} marked with \dag, no unique conclusion
could be drawn, although the linear fits indicate deviations.
With the clear deviation for choice \#7, $k=4$, and the predicition of the
Harris-Luck criterion in mind, also choice \#7 with $k=3$ might be
interpreted as a deviation from the Onsager value.
Choice \#5 and choice \#8 with $k=2$ are definitely compatible with
the Onsager exponent, but 
the deviations might just be too small to be detected.
In the random case, the fluctuations in $M$ were too large to even
confirm a power-law behaviour.  This is probably due to the random
fluctuations of the bond frequencies in the approximants used.
For choices \#1 to \#4, however, the picture was qualitatively the
same as in part (a) of \fref{fig:mass}.


\section{Summary and discussion}
\label{sec:summary}

We tested \IC\ for the Ising model on various periodic and aperiodic
planar graphs.
A procedure was described that makes use of special properties of the
algorithm in two dimensions and allows to determine the critical temperature 
with an accuracy of more than three significant digits.
The computational effort for this is small compared to other Monte
Carlo methods. 
Then we estimated the critical temperature on the twelvefold symmetric
square-triangle tiling and the tenfold symmetric T\"ubingen triangle
tiling, two cases for which no values were known before.
The values found are located in the rough range that one expects from
the first mean coordination number of the tilings.

In the second part of the paper, we presented a generalized version of
the algorithm applicable to models with non-identical coupling strengths.
We applied it to the Ising model on the Labyrinth tiling for three
different underlying substitutions, corresponding to irrelevant,
marginal, and relevant disorder according to the Harris-Luck
criterion, and a random case with relevant disorder. 
Each one was simulated for a few typical choices of the coupling strengths
on the exactly solvable submanifold, on the self-dual but not exactly
solvable submanifold, and away from both.  
The magnetic critical exponent $y_h=2-\beta/\nu$ was determined in all cases.
The values found were compatible with the Onsager value for all
self-dual subcases including the exactly solvable ones.  For the
latter this was known to be true exactly \cite{BGB}.
But it seems to extend to all self-dual cases even when the Harris-Luck
criterion predicts deviations.  
For some (but not all) other choices of the coupling strengths, when
the disorder was 
marginal or relevant according to the criterion, power-law behaviour with
different exponents was observed in the substitution systems.
The criterion can therefore be expected to be \emph{generically}
correct, but it does \emph{not} exclude exceptions on
lower-dimensional coupling manifolds.
The sharp contrast seen in the random version between self-dual and not
self-dual cases with respect to  the absence, respectively presence, of
large fluctuations of the mass of the cluster that wraps around the
system might
indicate that exceptions to the criterion are restricted to the
self-dual submanifold.  But further investigations are necessary to confirm
this observation.

The IC algorithm has proved a good tool for efficiently obtaining
quite accurate estimates of 
the critical temperature for periodic and aperiodic planar
graphs.  
Although only one of the two independent critical exponents, the
magnetic one, can be measured, this gives at least the possibility 
of detecting deviations from the Onsager universality class.  
Fortunately, it is a more sensitive quantity for this purpose than the
specific heat exponent $\alpha$, which is expected to remain zero when
the disorder is increased \cite{Dot}.

\ack

O. R. is indepted to Jon Machta for introducing him to the field of
cluster Monte Carlo algorithms during a nine-month stay in Amherst in
1996/7 and for important discussions during this work.
We thank Uwe Grimm and Joachim Hermisson for stimulating discussions.
Useful hints by the referees are gratefully acknowledged.
Financial support was given by the German Research Council (DFG).



\Bibliography{99}

\bibitem{Amm}
Ammann R, Gr\"unbaum B and Shephard G 1992
\arttit{Aperiodic tiles}
\textit{Discrete Comput.\ Geom.}\ \textbf{8} 1--25

\bibitem{MathQC}
Baake M 2001
\arttit{A guide to mathematical quasicrystals}
\emph{Quasicrystals} ed Suck J-B, Schreiber M and H\"au{\ss}ler P
(Berlin: Springer) to appear, preprint math-ph/9901014

\bibitem{BGB}
Baake M, Grimm U and Baxter R J 1994
\arttit{A critical Ising model on the Labyrinth}
\textit{Int.\ J. Mod.\ Phys.}\ \textbf{B8} 3579--600

\bibitem{Coord}
Baake M, Grimm U, Repetowicz P and Joseph D 1998
\arttit{Coordination sequences and critical points}
\textit{Proc.\ Int.\ Conf.\ Quasicrystals (Tokyo)} vol~6, 
ed Takeuchi S and Fujiwara T
(Singapore: World Scientific) pp 124--7, preprint cond-mat/9809110

\bibitem{STT}
Baake M, Klitzing R and Schlottmann M 1992
\arttit{Fractally shaped acceptance domains of quasiperiodic
square-triangle tilings with dodecagonal symmetry}
\textit{Physica} \textbf{A191} 554--8

\bibitem{TTT}
Baake M, Kramer P, Schlottmann M and Zeidler D 1990
\arttit{Planar patterns with fivefold symmetry as sections of periodic
structures in 4-space}
\textit{Int.\ J. Mod.\ Phys.}\ \textbf{B4} 2217--68

\bibitem{Bar}
Barber M N 1983
\arttit{Finite-size Scaling}
\textit{Phase Transitions and Critical Phenomena} vol~8,
ed Domb C and Lebowitz J L 
(London: Academic Press) pp 146--266

\bibitem{Bee}
Beenker F P M 1982 
\textit{Algebraic theory of non-periodic tilings of
the plane by two simple building blocks: a square and a rhombus}
Techn.\ Univ.\ Eindhoven, TH-report 82-WSK-04

\bibitem{Bha}
Bhattacharjee S M, Ho J-S and Johnson J A Y 1987
\arttit{Translational invariance in critical phenomena: Ising model on
a quasi-lattice}
\JPA \textbf{20} 4439--48

\bibitem{Bin}
Binder K 1981
\arttit{Finite size scaling analysis of Ising model block distribution
functions} 
\ZP \textbf{B43} 119--40

\bibitem{Bri}
Briggs K 1993
\arttit{Self-avoiding walks on quasilattices}
\textit{Int.\ J. Mod.\ Phys.}\ \textbf{B7} 1569--75

\bibitem{ICpar}
Choi Y S, Machta J, Tamayo P and Chayes L X 1999
\arttit{Parallel invaded cluster algorithm for the Ising model}
\textit{Int.\ J. Mod.\ Phys.}\ \textbf{C10} 1--16

\bibitem{Dot}
Dotsenko V 1995
\arttit{Critical phenomena and quenched disorder}
\textit{Usp.\ Fiz.\ Nauk.}\ \textbf{165} 287--96,
\textit{Physics Uspekhi} \textbf{38} 457--96

\bibitem{Fra}
Franzese G, Cataudella V and Coniglio A 1998
\arttit{Invaded cluster dynamics for frustrated models}
\PR \textbf{E57} 88--93

\bibitem{Aper}
Grimm U and Baake M 1997
\arttit{Aperiodic Ising models}
\textit{The Mathematics of Long-Range Aperiodic Order}
ed Moody R V (Dordrecht: Kluwer) pp 199--237

\bibitem{Har}
Harris A 1974
\arttit{Effect of random defects on the critical behaviour of Ising models}
\JPC \textbf{7} 1671--92

\bibitem{HerXY}
Hermisson J 2000
\arttit{Aperiodic and correlated disorder in XY-chains: exact results}
\JPA \textbf{33} 57--69

\bibitem{HerDiss}
Hermisson J 1999
\textit{Aperiodische Ordnung und magnetische Phasen\"uberg\"ange}
PhD thesis (Aachen: Shaker-Verlag)

\bibitem{HerICQ}
Hermisson J 1999
\arttit{Renormalization of two-dimensional Ising systems with
irrelevant, marginal and relevant aperiodic (dis)order}
\textit{Mat.\ Sci.\ Eng.}\ \textbf{A} to appear

\bibitem{HerQC}
Hermisson J, Grimm U and Baake M 1997
\arttit{Aperiodic Ising quantum chains}
\JPA \textbf{30} 7315--35

\bibitem{Led}
Ledue D, Landau D P and Teillet J 1995
\arttit{Static critical behavior of the ferromagnetic Ising model on
the quasiperiodic octagonal tiling}
\PR \textbf{B51} 15523--30

\bibitem{Luc}
Luck J M 1993
\arttit{A classification of critical phenomena on quasi-crystals and
other aperiodic structures}
\textit{Europhys.\ Lett.}\ 359--64

\bibitem{Mac}
Machta J 1998 Private communication

\bibitem{IC1}
Machta J, Choi Y S, Lucke A, Schweizer T and Chayes L M 1995
\arttit{Invaded cluster algorithm for equilibrium critical points}
\PRL \textbf{75} 2792--5

\bibitem{IC2}
Machta J, Choi Y S, Lucke A, Schweizer T and Chayes L M 1996
\arttit{Invaded cluster algorithm for Potts models}
\PR \textbf{E54} 1332--45

\bibitem{ICdyn}
Moriarty K, Machta J and Chayes L Y 1999
\arttit{Dynamic and static properties of the invaded cluster algorithm}
\PR \textbf{E59} 1425--34

\bibitem{Oka}
Okabe Y and Niizeki K 1988
\arttit{Monte Carlo simulation of the Ising model on the Penrose lattice}
\JPSJ \textbf{57} 16--9

\bibitem{Pen}
Penrose R 1974 
\arttit{The r\^ole of aesthetics in pure and applied mathematical research}
\textit{Bull.\ Int.\ Math.\ Appl.}\ \textbf{10} 266--71

\bibitem{Red}
Redner O 1999 
\textit{Effiziente Simulation periodischer und nicht-periodischer
Ising-Modelle am kri\-ti\-schen Punkt} 
diploma thesis, available from the author

\bibitem{HighTemp}
Repetowicz P, Grimm U and Schreiber M 1999 
\arttit{High-temperature expansion for Ising models on quasiperiodic tilings}
\JPA \textbf{32} 4397--418

\bibitem{KacWard}
Repetowicz P, Grimm U and Schreiber M 1999 
\arttit{Planar quasiperiodic Ising models}
\textit{Mat.\ Sci.\ Eng.}\ \textbf{A} to appear,
preprint cond-mat/9908088

\bibitem{Sir}
Sire C, Mosseri R and Sadoc J-F 1989
\arttit{Geometric study of a 2D tiling related to the octagonal
quasiperiodic tiling} 
\textit{J. Phys.\ France} \textbf{50} 3463--76

\bibitem{Soe}
S{\o}rensen E S, Jari\'c M V and Ronchetti M 1991
\arttit{Ising model on Penrose lattices: boundary conditions}
\PR \textbf{B44} 9271--82

\bibitem{SwWa}
Swendsen R H and Wang J-S 1987
\arttit{Nonuniversal critical dynamics in Monte Carlo simulations}
\PRL \textbf{58} 86--8

\bibitem{Syo}
Syozi I 1972
\arttit{Transformation of Ising models}
\textit{Phase Transitions and Critical Phenomena} vol~1,
ed Domb C and Green M S (London: Academic Press) pp 269--329

\bibitem{Wol}
Wolff U 1989
\arttit{Collective Monte Carlo updating for spin systems}
\PRL \textbf{62} 361--4


\endbib
\end{document}